%% file: 0_arxiv_version_main.tex
\documentclass[sigconf]{acmart}
\AtBeginDocument{%
  }

\usepackage{ulem}

\input{includes/packages.tex}

\input{includes/macros.tex}

\usepackage{float}

\begin{document}

\title{CSRec: Rethinking Sequential Recommendation from A Causal Perspective.}

\author{Xiaoyu Liu}
\authornote{Both authors contributed equally to this research.}
\email{xliu1231@umd.edu}
\affiliation{%
  \institution{University of Maryland, College Park}
  \city{}
  \country{}}

\author{Jiaxin Yuan}
\authornotemark[1]
\email{jyuan98@umd.edu}
\affiliation{%
  \institution{University of Maryland, College Park}
  \city{}
  \country{}}

\author{Yuhang Zhou}
\email{tonyzhou@umd.edu}
\affiliation{%
  \institution{University of Maryland, College Park}
  \city{}
  \country{}}
  
\author{Jingling Li}
\email{jinglingli1024@gmail.com}
\affiliation{%
  \institution{University of Maryland, College Park}
  \city{}
  \country{}}
  
\author{Furong Huang}
\email{furongh@umd.edu}
\affiliation{%
  \institution{University of Maryland, College Park}
  \city{}
  \country{}}
  
\author{Wei Ai}
\email{aiwei@umd.edu}
\affiliation{%
  \institution{University of Maryland, College Park}
  \city{}
  \country{}}

\renewcommand{\shortauthors}{Liu et al.}

\begin{abstract}
 
The essence of sequential recommender systems (RecSys) lies in understanding how users make decisions. Most existing approaches frame the task as sequential prediction based on users' historical purchase records. While effective in capturing users' natural preferences, this formulation falls short in accurately modeling actual recommendation scenarios, particularly in accounting for how unsuccessful recommendations influence future purchases. Furthermore, the impact of the RecSys itself on users' decisions has not been appropriately isolated and quantitatively analyzed.
To address these challenges, we propose a novel formulation of sequential recommendation, termed Causal Sequential Recommendation (CSRec). Instead of predicting the next item in the sequence, CSRec aims to predict the probability of a recommended item's acceptance within a sequential context and backtrack how current decisions are made. Critically, CSRec facilitates the isolation of various factors that affect users' final decisions, especially the influence of the recommender system itself, thereby opening new avenues for the design of recommender systems. CSRec can be seamlessly integrated into existing methodologies. Experimental evaluations on both synthetic and real-world datasets demonstrate that the proposed implementation significantly improves upon state-of-the-art baselines. {\color{blue}\href{https://github.com/margotyjx/CSRec_repo.git}{github link here.}}

\end{abstract}





\maketitle

\input{sections/1_Introduction}

\input{sections/2_Preliminary}

\input{sections/3_Causal_Graph_of_Sequential_Recommendation}

\input{sections/4_Causal_Formulation}

\input{sections/5_Estimate_System_Effect}

\input{sections/Experiment}

\input{sections/Related_Work}

\input{sections/Conclusion}


\clearpage

\bibliographystyle{ACM-Reference-Format}
\bibliography{reference}

\appendix

\input{sections/app3_prelim}

\input{sections/app1_thm}

\input{sections/app2_exp}

\end{document}

%% file: includes/packages.tex
\usepackage[textsize=tiny]{todonotes}

\usepackage{amsmath,amsfonts,amsthm}
\usepackage{mathtools}
\usepackage{bm}
\usepackage{subcaption}
\usepackage{cleveref}
\usepackage{xspace}
\usepackage{wrapfig, framed, caption}
\usepackage{makecell}

\usepackage{extarrows}

\usepackage{enumitem,kantlipsum}
\usepackage{thm-restate}

\usepackage{xcolor}
\usepackage{hyperref}
\hypersetup{
    linktoc=all,
    colorlinks=true,
    citecolor=green,
    filecolor=black,
    linkcolor=blue,
    urlcolor=blue,
}

\usepackage{xspace}
\usepackage{multirow}

%% file: includes/macros.tex
\newcommand{\CSRec}{CSRec\xspace}
\newcommand{\MSSD}{MSSD\xspace}

\DeclareMathOperator*{\argmax}{arg\,max}

\newtheorem{theorem}{Theorem}[section]

\theoremstyle{definition}
\newtheorem{definition}[theorem]{Definition}

\newtheorem{rules}[theorem]{Rule}
\theoremstyle{remark}
\newtheorem{remark}[theorem]{Remark}

\newcommand{\indep}{\perp \!\!\! \perp}

\newcommand{\defeq}{\vcentcolon=}

%% file: sections/1_Introduction.tex
\section{Introduction}
\label{sec:intro}


In the contemporary data-driven landscape, the significance of building a recommender system (recsys) cannot be overstated. These systems form the foundation for delivering personalized user experiences, improving customer satisfaction, and optimizing decision-making processes across a wide range of industries.

The fundamental objective of recommender systems is to understand how users make decisions and to provide support in this process. Most existing works, however, equate this objective with recommending the item that users are likely to prefer the most. Consequently, these approaches focus on estimating users' inherent preferences based on observational data, such as purchasing records, click data, and item reviews \cite{li2024recent}, and develop various methods to mitigate biases, such as popularity bias and exposure bias \cite{klimashevskaia2024survey, gao2024causal}.
Later, the study of sequential recommendation recognizes that users' preferences are also affected by temporal factors, taking into consideration that preferences may evolve over time or that previous purchase decisions can influence future choices. Consequently, recommendation tasks are formulated to predict the most likely next purchased item based on previous purchases, offering a more personalized recommendation \cite{wang2019sequential,tang2018caser,shin2024BSARec}.

However, recovering users' preferences is not fully equivalent to understanding the decision-making process. Despite their notable success, such formulations tend to oversimplify the complexities of real-world recommendation scenarios and fall short in precisely characterizing the recommendation process. This oversimplification, to some extent, limits the potential for developing truly effective recommender systems.

Recommendation differs from users' natural selection processes in that the recsys controls how items are exposed to users, effectively serving as an \textit{intervention} in their purchase decisions. In the formulation of predicting the next possible purchase, predictions are typically based on the user's previous acceptance of a series of items. However, it is entirely plausible that a user may dislike the system's recommendations or show no interest in the suggested items, leading to sequences that include both acceptance and rejection of items. Unfortunately, these possible sequences are excluded from current formulations\cite{Zhou2022FMLPRec, shin2024BSARec, klimashevskaia2024survey} due to the fundamental difference in the training goal.

In addition, an increasing number of recent works have recognized the interventional nature of recommender systems and have attempted to model the system's influence on users. For instance, \cite{ge2020understanding} highlighted the phenomenon of echo chambers, where users receive increasingly homogenized recommendations. Similarly, \cite{kalimeris2021preference} sought to address the issue of amplified preference loops, where user preferences in observational data are repeatedly reinforced, ultimately undermining the personalized recommendation experience. The emergence of these problems underscores the need for an analytic framework capable of measuring the system's influence on users in a differentiable manner. While specific solutions have been proposed, they often overlook the fact that users' decisions are collaboratively influenced by multiple factors, not just the system itself. This reveals a gap in the existing methods, which lack the ability to isolate the system's influence from other factors, such as users' inherent preferences or previous purchase decisions.

The limitations of current next-item prediction formulations, coupled with the absence of a comprehensive analytical framework for understanding how users make decisions, highlight a critical gap in the existing research on recommender systems. To address this gap, it is essential to adopt a perspective that can rigorously characterize the complex causal relationships between the various factors influencing user decisions. This is where causal inference offers a unique and powerful approach.

In this paper, we introduce a novel formulation called CSRec (\textbf{C}ausal \textbf{S}equential \textbf{R}ecommendation), which can be integrated into existing recommender systems to preserve the underlying causal structure in sequential recommendations, thereby overcoming existing limitations.
Given a causal graph, the causal inference perspective offers an explicit and differentiable framework to isolate and quantify the influence of each variable on a user's decision. While the underlying causal graph may differ across contexts, the same procedure can be consistently applied, making it a flexible and broadly applicable framework.

\input{tex/causal_graph}
In this paper, we resort to a general causal graph to model the sequential recommendation scenario, where the system's recommendations, the user's past purchases, and the user's preferences collaboratively influence the user's decisions on potential items. By analyzing this causal graph, we identify how a user's decision at the current step is influenced by previous decisions, the user's preferences, and the system's recommendations, conceptualized as a weighted sum over different possible outcomes (acceptance or rejection).

Moreover, we introduce a method to quantitatively analyze the influence of recommender systems on users through treatment effect estimation, defined as the average difference in a user's decisions before and after the recommendation. This formulation opens up new possibilities for developing personalized recommender systems. In Section \ref{sec:treatment-effect}, we present several examples demonstrating how this approach can be used as an evaluation metric to assess whether users are building trust in the system. Additionally, it offers a framework for developing strategies to balance the exploration-exploitation trade-off when navigating users' preferences in unknown domains.

We conduct extensive experiments on both synthetic and real-world dataset and empirically show that this simple adjustment of existing baselines generally improves the performance and outperforms SOTA baselines.

\textbf{Summary of contributions:} \\
\textbf{(1)} We identify the inherent difference between the processes of recommendations and users' natural selection, highlighting that existing formulations fail to account for cases where users' decisions are adversely affected by unsuccessful recommendations. \\
\textbf{(2)} To address this gap, we propose Causal Sequential Recommendation (CSRec). CSRec is the first analytical framework that explicitly considers how various factors causally and collaboratively influence users' decisions. Additionally, it can isolate the impact of individual factors, such as the recommender system itself, on users' decisions, thereby opening up new possibilities in the design and development of recommender systems. \\
\textbf{(3)} CSRec is a general framework that can be applied to a wide range of recommendation scenarios and can be easily integrated into existing recommender systems for improvement. We conducted extensive experiments and case studies using both synthetic and real-world datasets, demonstrating competitive results across various baselines.

%% file: tex/causal_graph.tex
\begin{figure*}[!ht]
    \centering
    \includegraphics[width=\textwidth]{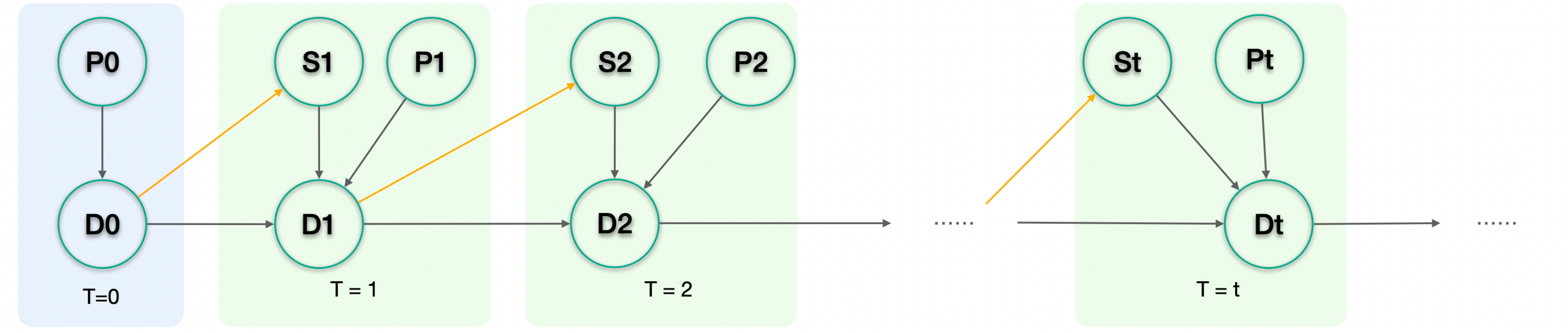}
    \caption{\textbf{Causal Graph for Sequential Recommendation.}}
    \label{fig:causal-graph}
\end{figure*}

%% file: sections/2_Preliminary.tex
\section{Background and Preliminary}
\label{sec:prelim}

The goal of training a recommender system varies across different contexts, such as maximizing the likelihood of a recommended item being accepted or enhancing long-term trust and user satisfaction. At the heart of developing any recommender system is the need to understand how users make decisions and how these decisions can be influenced.

While existing research has made significant strides in recovering users' natural preferences from observational data, two critical aspects are under-emphasized in sequential recommendation: (1) \textbf{System-Controlled Exposure}: Unlike users' spontaneous purchases, recommender systems control how items are exposed to users. It is not uncommon for users' purchase decisions to be adversely affected by system recommendations. However, this influence is often overlooked by existing formulations due to differences in the underlying training goals. (2) \textbf{Collaborative Influence on Decision-Making}: Users' decisions are shaped by various recommendation scenarios and design needs, necessitating a unified and systematic approach to analyze, understand, and evaluate these influences.

In this section, we begin by reviewing how existing methods approach recommendation. We then reformulate the recommendation process from a causal perspective, introducing the basic concepts of causal inference and demonstrating how to evaluate causal relationships using a causal graph. Finally, we apply these methods to analyze and model the causal relationships within subsequent purchases, as discussed in \cref{sec:causal-graph} and \cref{sec:constraint}.

\noindent \textbf{Recommendation as Sequential Prediction: The Traditional Viewpoint.}
Sequential recommender systems model historical user interactions as temporally-ordered sequences to recommend potential items that users are interested in. In the general formulation as a sequential dataset, we are given an item set $\mathcal{I}$ and a user's interaction sequence $s = \{i_1, i_2, ..., i_n\}$, $i \in \mathcal{I}$ in temporal order where $n$ is the length of the sequence\cite{li2023text}. This formulation only accounts for cases where all items in the sequence are accepted by users. However, during the recommendation process, users may also be recommended items in which they are not interested. In other words, if we distinguish between exposure and acceptance, with acceptance denoted as 1 and rejection as 0, the traditional viewpoint only models sequences composed entirely of 1s, thereby overlooking sequences that include 0s.

\noindent \textbf{Recommendation from a causal perspective.}
Compared to the existing formulation that focuses on uncovering users' unbiased interests, causal inference methods offer a more general and intuitive approach by aiming to characterize how users' decisions are causally influenced. In this context, the recommender system acts as an \textit{intervention} on the user, specifically by controlling the exposure of items to the user, which naturally separates the concepts of exposure and acceptance.

\noindent \textbf{Causal graph through DAG.} The causal relationships among variables can be represented by a Directed Acyclic Graph (DAG). In this graph, each (potentially high-dimensional) variable is represented as a node, and the directed edges indicate causal relationships, always pointing from parent nodes to child nodes. A causal graph can be estimated using do-calculus  as detailed in \Cref{app: do-calculus}.

\noindent \textbf{Intervention and do-operator.} \textit{Intervention} is one of the fundamental concepts in causal inference. When we intervene on a variable, we set the value of a variable and cut all incoming causal arrows since its value is thereby determined only by the intervention~\cite{pearl2012calculus}. The intervention is mathematically represented by the \textit{do-operator} $do(\cdot)$. We include the do-calculus for identifying causal effects from a causal graph in \cref{app: do-calculus}.
In the context of recommendation, we are trying to understand how the \textit{changes in recommended items affect the user's decision and satisfaction}, which can be represented by 
\begin{definition}[Recommendation in Causal Inference]
    \label{def:causal-rec}
    Let $D$ and $S$ represents the user decision and the system recommendation respectively, 
    \begin{equation}
        \label{eq:causal-rec}
        P(D | do (S=s), P)
    \end{equation}  
    characterizes an interventional distribution and reflects how a change of $S$ affects the distribution of $D$ for users represented by the preference $P$ . 
\end{definition}

\noindent \textbf{Observational distribution and interventional distribution.}
Observational distributions refer to data passively collected without any explicit manipulation, reflecting correlations influenced by potential underlying factors. In contrast, interventional distributions are generated through controlled interventions, where specific variables are deliberately manipulated to observe causal outcomes. In this paper, since we are interested in exploring how system intervention affect user's purchase decision, we refer to historical data,  where user selects what to buy, as the observational data, and refer to the data which contains user's reaction towards recommender system as the interventional data.

\noindent\textbf{Treatment effect estimation.}
Treatment effect estimation is a fundamental aspect of causal inference that involves quantifying the impact of an intervention on an outcome. It measures the difference in outcomes resulting from a specific treatment compared to what would have occurred in the absence of the treatment. This method offers a rigorous approach to evaluating the influence of recommender systems on user decisions. By assessing the variance in user decisions before and after receiving recommendations, we can accurately quantify the system's impact on user behavior.

%% file: sections/3_Causal_Graph_of_Sequential_Recommendation.tex
\section{Sequential Recommendation Represented in Causal Graph} 
\label{sec:causal-graph}

In this section, we introduce the causal graph for the sequential recommendation as shown in \Cref{fig:causal-graph}. 
Without the intervention of recommendation system, users select which item to purchase according to their own natural preference, and this purchase/rating/clicking behavior is collected as the observational/historical data. We use $P$ to denote the user's preference. Since the sequential nature of the data, we use subscript to denote the timestamp. Thus $P_t$ reflects the user's preference at time step $t$. We use $S$ to denote the exposure of items to users. Specifically, under an interventional setting, it represents the system's recommendation and under an observational setting, it means that users are exposed to the item from a different source. We use $D$ to denote user's decision towards inquired items, and $D$ to denote the user's decision towards the recommended items.

In this paper, we focus on one of the most general recommendation scenarios, where a user's decisions are influenced by their inherent preferences, the recommender system, and their previous decisions. While in an online setting, there might be an additional causal link from $D_t$ to $S_{t+1}$, in this work, we adhere to an offline training schedule and do not consider the causal relationship depicted by the yellow arrow.


This causal graph representation provides a way to overcome the limitations of existing formulations while retaining the benefits of recovering users' preferences. In this context, $P_t$ represents the user's genuine interests in the items, which can still be modeled by existing methods. The key difference is that, rather than identifying the item the user likes the most (item with the highest probability of $P_t$), we focus on finding the item the user is most likely to purchase (item with the highest probability of $D_t$). Such a shift allows for the natural consideration of all influential factors and the underlying causal relationships, bringing the approach closer to the original goal of understanding the user's decision-making process.

%% file: sections/4_Causal_Formulation.tex
\section{Demystifying User's Decision-Making Through Causal Formulation}
\label{sec:constraint}

In this section, we demonstrate how to estimate the impact of various factors on users' decisions. Specifically, we adhere to the causal graph outlined in \Cref{fig:causal-graph} and illustrate how users' decisions are influenced by the recommender system, previous decisions, and their inherent preferences.

\subsection{Problem Setting}
From a causal perspective, estimating how a sequence of system recommendation affects user's decision at $T=t$ is equivalent to estimating the causal effect of the decision under intervention from the system.
\begin{equation}
    \label{eq: estimate-sym-user}
    \mathbb{P}(D_t | \text{do}(S_t, S_{t-1},...,S_1), P_t).
\end{equation}
We intervene only in system recommendations to improve the purchasing likelihood under the recommendations. We cannot directly change a user's behavior or purchasing logic; instead, we aim to increase the probability of purchase through our recommendations.

From \Cref{fig:causal-graph}, we observe that the decision $D_t$ is directly influenced by the recommender system $S_t$ at time $t$, the user's preference $P_t$, and the user's previous decision $D_{t-1}$. In a sequential recommendation setting, $D_{t-1}$ inherently includes information about the recommender system at the prior time step $S_{t-1}$, the user's preference $P_{t-1}$, and the decision $D_{t-2}$. Therefore, while the most recent decision is a direct cause, the entire history can indirectly affect the user's current decision.

\subsection{Estimating the Interventional Distribution.}
To estimate such an interventional distribution, we need to (1) sequentially recover the causal relationship at the current timestamp and (2) maintain such causal structure in the training of a recommender system. We start with revealing the causal dependency between two subsequent decisions under a series system recommendations in the following theorem, which naturally reveals a constraint to maintain the causal structure. 

\begin{theorem}
\label{thm:prob_dist}
    At time $T = t$, the probability distribution for a user's decision given precedent recommendations $\text{do}(S_{t-1},...,S_1)$ and the preference $P_t$ follows the equality 
    \begin{multline}
        \mathbb{P}(D_t | \text{do}(S_t, S_{t-1},...,S_1), P_t) = \sum_{D_{t-1}}
        \mathbb{P}(D_t | S_t, D_{t-1}, P_t)\cdot \\
        \mathbb{P}(D_{t-1} | \text{do}(S_{t-1},...,S_1), P_t) 
    \end{multline}
\end{theorem}

The proof for the theorem is shown in \Cref{app:proof_thm}.

\Cref{thm:prob_dist} describes how user's current decision is affected by previous decisions, the system and the user's preference. To explain this theorem in details, we write this summation for time $t = 2$, denoting the system recommended items as $s_1, s_2$ for $t = 1$ and $t = 2$ respectively. 
For simplicity, we consider $D_{t} \in \{0,1\}$ where $0$ denotes a negative decision and $1$ denotes a positive decision. 
We can write
\begin{align*}
\small
    & \mathbb{P}(D_2 | \text{do}(S_2 = s_2, S_1 = s_1), P_2)  \\
 = & \sum_{D_1\in \{0,1\}} \mathbb{P}(D_2 | S_2 = s_2, D_1,, P_1 )\mathbb{P}(D_1 | \text{do}(S_1 = s_1), P_1)\\
     = & ~ \mathbb{P}(D_2 | D_1 = 1 ,S_2 = s_2, P_2)\mathbb{P}(D_1 = 1 | \text{do}(S_1 = s_1), P_1) + \\
    & \quad \space \mathbb{P}(D_2 | D_1 = 0 ,S_2 = s_2, P_2)\mathbb{P}(D_1 = 0 | \text{do}(S_1 = s_1), P_1).
\end{align*}
Here $do(\cdot)$ represents the interventional distribution where the recommender system controls the exposure as stated in \Cref{sec:prelim}, and $\mathbb{P}(D_2 | D_1 = 1 ,S_2 = s_2, P_2)$ represents the observational distribution such that the source of exposure is not the recommender system.

Hence, the estimation of this probability is equivalent to the prediction of one user's reaction to the recommended item $s_2$, taking their decision on $s_1$ into consideration, and weighted by the corresponding interventional probability. 
So long as this equality maintained at each timestamp, the causal relationship remains intact throughout the entire recommendation process.

\subsection{Maximizing Users' Positive Decisions.}

A general goal of a recommender system is to maximize the probability of users' positive decisions, e.g., purchase or click rate. Existing works equate such a goal as selecting the item that has the highest preference score for users. We on the other hand, consider to maximize the purchasing probability by adjusting system's behavior. This subtle distinction enables us to first generalize the concept and then incorporate \Cref{thm:prob_dist} to maintain the inherent causal structure.

Instead of examining the user's preference towards each product, we compute the probability of positive decisions given a series of recommendations $\text{do}(S_t, S_{t-1},...,S_1)$, together with their current preference $P_t$. Hence, we can formulate the task at time $t$ as an optimization problem: 
\begin{equation}
\label{eq: optimization0}
    s_t = \argmax_{\text{do}(S_t) \in \mathcal{S}} \mathbb{P}(D_t | \text{do}(S_t, S_{t-1},...,S_1), P_t).
\end{equation}
Note that an alternative way to define the optimization problem is to consider the cumulative decisions $\sum_{i = 1}^T D_i$. For simplicity we consider a one-step maximization as \Cref{eq: optimization0} only, but the framework can be generalized to cumulative decisions easily.

To preserve the causal structure, we refer to \Cref{thm:prob_dist} and accordingly propose CSRec (\textbf{C}ausal \textbf{S}equential \textbf{Rec}ommender) with this new optimization goal as the following: 
\begin{equation}
    \label{eq: optimization1}
    \max_{\text{do}(S_t) \in \mathcal{S}} \mathbb{P}(D_t | \text{do}(S_t,...,S_1), P_t)
\end{equation}
\begin{multline}
    \label{eq:constraint}
    \text{s.t.} \quad \mathbb{P}(D_t | \text{do}(S_t,...,S_1), P_t) = \sum_{D_{t-1}} \mathbb{P}(D_t | S_t, D_{t-1}, P_t) \cdot \\
    \mathbb{P}(D_{t-1} | \text{do}(S_{t-1},...,S_1), P_t) 
\end{multline}

\begin{remark}
\label{rmk: extension}
Concretely, CSRec considers optimizing the probability of users' positive decision rate by incorporating the causal relationship for a variety of factors.
\end{remark}

\subsection{Enforcing the constraint.}
To enforce the constraint \Cref{eq:constraint}, we need to estimate quantities from both the interventional distribution $\mathbb{P}(D_{t-1} | \text{do}(S_{t-1},...,S_1), P_t)$ and observational distribution $\mathbb{P}(D_t | S_t, D_{t-1}, P_t)$. Here we emphasize the difference between these two distributions. The $do(\cdot)$ notation in $\mathbb{P}(D_{t-1} | \text{do}(S_{t-1},...,S_1), P_t)$ refers to situations where the exposure is completely controlled by recommendation while $S_t$ in observational distribution $\mathbb{P}(D_t | S_t, D_{t-1}, P_t)$, where $S_t$ represents to items exposed to users from a different source than a recommender system.

Hence, probability $\mathbb{P}(D_t | S_t, D_{t-1}, P_t)$ can be understood as users' natural decision behavior $D_t$ towards items $S_t$ based on their previous decision $D_{t-1}$ and temporal preference $P_t$. This is a natural goal for a traditional neural network based recommendation system such as SASRec \cite{SASRec2018} or BSARec \cite{shin2024BSARec}. Therefore, we estimate this quantity using an neural network recommender system trained on observational dataset. 

Putting everything together, we build a recommender system that models users' decisions under the intervention of recommendations, parameterized as $f_t(\theta) := \mathbb{P}(D_t | \text{do}(S_t,...,S_1), P_t)$. Consequently, \Cref{eq:constraint} can be written as
%
\begin{equation}
   f_t = \sum_{D_{t-1}} \mathbb{P}(D_t | S_t, D_{t-1}, P_t)f_{t-1}.
\end{equation}
%
We adapt a sequential recommender system trained on historical data, and parameterize it as $\Tilde{f}$:
%
\begin{equation}
     \mathbb{P}(\cdot |S_t, D_{t-1}, P_t) = \Tilde{f}(D_{t-1},...,D;\eta). 
\end{equation}
%
%
With the general objective function of existing sequential recommender systems, we trained our model using the following objective
\begin{multline}
    \label{eq:loss fn 1}
    \mathcal{L}_t = -(D_t \log f_t(\theta) + (1-D_t) \log (1-f_t(\theta))), \quad \\
    \text{s.t. } f_t(\theta) = \sum_{D_{t-1}} f_{t-1}(\theta) \Tilde{f}(\eta).
\end{multline}
By the Lagrangian multiplier method, we write the loss function for optimizing $f_t(\theta)$ as
\begin{multline}
    \label{eq:loss fn 2}
    \mathcal{L}_t = -(D_t \log f_t(\theta) + (1-D_t) \log (1-f_t(\theta))) + \\
    \left\| f_t(\theta) - \sum_{D_{t-1}} f_{t-1}(\theta) \Tilde{f}(\eta)\right\|_2.
\end{multline}
In summary, we first train a pre-selected sequential recommender system on the observational dataset and use that to estimate the quantity $\Tilde{f} = \mathbb{P}(\cdot |S_t, D_{t-1}, P_t)$. Then we trained our model on interventional dataset to estimate $\mathbb{P}(D_{t-1} | \text{do}(S_{t-1},...,S_1), P_t)$ using the objective function \Cref{eq:loss fn 2}.

We note here that we can use any recommender system to estimate $\Tilde{f} = \mathbb{P}(\cdot |S_t, D_{t-1}, P_t)$ so long as its goal is to estimate users' natural decision behavior. We show in \Cref{sec: exp} the performance of CSRec integrated with four different sequential recommender systems.

%% file: sections/5_Estimate_System_Effect.tex
\section{Estimate System's Influence Through Treatment Effect}
\label{sec:treatment-effect}

With the decision-making process explicitly characterized as shown in \Cref{sec:constraint}, we can approximate the potential outcome for each candidate item to recommend, specifically the probability that the recommendation will be accepted.

Beyond approximating the user's decision process, it is also essential to isolate and quantify the system's influence on its users. For example, in the exploration and exploitation dilemma, people must predict the risk of recommending a novel item that is out-of-the-distribution. Additionally, when a system is trying to gather information about a new user, it is crucial to quickly understand the user without over-exploiting their trust in the system. Quantifying the system's influence on users can significantly enhance the development of a personalized and trustworthy system, moving beyond merely predicting user preferences from static data. Estimating how one variable affects another is addressed by measuring the \textbf{treatment effect} in causal inference.

\subsection{Treatment effect in recommendation}

Treatment in causal inference refers to the action that applies (exposes, or subjects) to a unit\cite{pearl2012calculus}. For each unit-treatment pair, the outcome of the treatment when applied to the unit is referred as the potential outcome. With N treatments $T = \{1,2,3,..., N\}$, the potential outcome of applying treatment $T_i$ is denoted as $Y(T=T_i)$.
The treatment effect can be quantitatively defined as follow.
\begin{definition}[Binary Average Treatment Effect(ATE)] Suppose we want to measure the treatment effect of a treatment $T=1$. Then the average treatment effect is defined as:
\begin{equation}
    \label{def:ate}
    \mathbb{E}[Y(T=1) - Y(T=0)]
\end{equation}
where $Y(T=1)$ and $Y(T=0)$ denote the potential treated and control outcome of the whole population respectively.
\end{definition}
In the recommendation scenario, we extend \Cref{def:ate} to the difference between the user's decision with and without the recommendation.

\begin{definition}[Treatment Effect in Recommendation]
    \label{def: ate-in-rec}
    \begin{equation}
     \text{TER}_t \defeq \mathbb{P}(D_t | \text{do}(S_t, S_{t-1},...,S_1), P_t) -  \mathbb{P}(D_t | S_t, S_{t-1},...,S_1, P_t).
\end{equation}
\end{definition}

Here $S_t$ represents the item exposed to users at timestamp $t$. In the observational setting ($\mathbb{P}(D_t | S_t, S_{t-1},...,S_1, P_t)$) the exposure is controlled by users or randomized, while in the interventional settings ($\mathbb{P}(D_t | \text{do}(S_t, S_{t-1},...,S_1), P_t)$), it is controlled by the recommender system.

We compute the first term using our proposed model CSRec and the second term using any commonly used recommender systems: 
\begin{equation}
    \text{TER}_t(V) = f_t(V, X_{\text{intv}}; \theta) - \Tilde{f}(V, X_{\text{obs}}; \eta)
\end{equation}
where $V$ denotes the set of all items, $X_{\text{intv}}$ and $X_{\text{obs}}$ represent interventional data and observational data respectively.

We then demonstrate two cases where the quantified system's influence on users effectively broadens the perspective of building a recommendation system. 

\paragraph{Evaluating Recommender System}

Quantifying the system's influence in a causal perspective could provide more insight into the effectiveness of the recommender system. It allows for understanding how recommendations change and shape users' decisions over time, beyond the single accuracy metrics. Moreover, by inspecting the treatment effect over different users (heterogeneous treatment effect), we can probe the mechanism of the recommender system and increase the interpretability. 

\paragraph{Exploration and Exploitation}

The quantified system's influence on each timestamp could also contribute to building a balanced recommender system between exploration and exploitation. The system could more precisely identify which time to give the suggestions based on the existing knowledge (exploitation) and when to test the new items (exploration). This dynamic strategy accommodates the overall user experience and allows the discovery of new content. 

%% file: sections/Experiment.tex
\section{Experiment}
\label{sec: exp}
\input{tex/gpt4_table}

In this section, we incorporate CSRec into various sota baselines \cite{hidasi2016GRU4Rec, tang2018caser, SASRec2018, shin2024attentive, sun2019bert4rec, chang2019sequential} and compare the performance on both synthetic and real-world datasets. 
 
We aim to answer the following research questions:

\indent \textbf{RQ1.} How does the proposed method perform on interventional data compared to baseline sequential recommendation (SR) models?\\
\indent \textbf{RQ2.} Would the proposed method degrade performance on observational data?\\
\indent \textbf{RQ3.} How does the proposed scheme perform when integrating with different existing recommendation systems?

In the following section, we will describe the experimental settings and answer the above research questions.

\subsection{Basic setup}
\label{sec:exp_setup}
\subsubsection{Datasets}
we evaluate \CSRec on two datasets: a real-world Spotify Sequential Music dataset\cite{brost2019music}, and a synthetic book dataset, GPT4-books where the generation details can be found in \Cref{app: exp-details}. \MSSD was first introduced at the WSDM Cup 2019 Spotify Sequential Skip Prediction Challenge \cite{brost2019music}. The dataset contains approximately 160 million listening sessions, associated user actions, and 3.7 million tracks with acoustic features. The task is to predict whether a particular user will skip individual tracks encountered in a listening session. We use sessions sampled from personalized recommendations and user's own collections.

Given the limited access to interventional data, where the recommender system fully controls the exposure, we also created a synthetic dataset using GPT-4 as agents to mimic users' reactions to certain recommendations. We use the Amazon book dataset\cite{ni2019justifying} as a list of potential items to recommend. Then, we prepare 10000 users, each with preferences across 10 genres. Detailed information can be found in \Cref{app: data-generation}.

\subsubsection{Baselines and tasks}
We compare CSRec with various sota baselines with different backbones: (1) RNN or CNN-based sequential models: GRU4Rec \cite{hidasi2016GRU4Rec} and Caser \cite{tang2018caser} and (2) transformer-based sequential recommender systems: SASRec \cite{SASRec2018}, BERT4Rec \cite{sun2019bert4rec}, FMLPRec \cite{Zhou2022FMLPRec}, and BSARec \cite{shin2024BSARec}. 
We further show the performance of the proposed methods when integrated with existing recommendation models on synthetic data GPT4-books. Specifically, we integrate CSRec with SASRec, BERT4Rec, FMLPRec and BSARec.
In our large-scale experiment on MSSD, in addition to the aforementioned baselines, we also include a performance comparison before and after incorporating CSRec into the WSDM Cup 2019 winning model, Few-Shot \cite{chang2019sequential}, as it has specialized optimization towards MSSD.

\subsubsection{Evaluation metrics}
\vspace{-.5em}
We evaluate CSRec and the baselines models on both observational data where exposure to items does not come from recommendations and the interventional data where recommender system controls the exposure. 
On the observational data, we compare their performance in the task of forecasting the user's subsequent action from prior actions. 
While on the interventional data, we estimate the user's decision regarding a recommendation. Given the difference in the objective that our model and other recommender system aim to achieve, necessary adaptation is required and we provide the detailed procedure as follows.

\noindent\textbf{Observational data.}
On observational data, we compare CSRec and SR baselines via widely used Top-k metrics, HR@k (Hit Rate) and NDCG@k (Normalized Discounted Cumulative Gain) where k is set to 5, 10, and 20 \cite{metrics_2020}. HR@k measures whether the ground-truth item is in the top-k positions of the recommendation list. NDCG on the other hand, takes the rank into considerations that it assigns higher scores to higher positions in the recommendation list. 

We would like to point out a natural adjustment we make on CSRec for evaluation on observational data here. Even though CSRec is configured to model $\mathbb{P}(D_t | \text{do}(S_t,...,S_1), P_t)$ under the interventional setting, on historical data, we can safely assume that all previous purchased items were recommended and accepted. The output probabilities are then ranked and evaluated using HR@k and NDCG@k.

\input{tex/mssd_table}

\noindent\textbf{Interventional data: }
On interventional data, the dataset contains two sequences: recommended items and users' corresponding decisions $\in \{0,1\}$ indexed with time. We aim to examine how well models estimate user's decision when being recommended on interventional data. 

For CSRec, the model outputs the predicted users' purchase probability for the next recommended items. We then compare this resulting probability with the ground truth decisions via an adjusted hit rate defined as follows
\begin{equation}
    \text{AHR}_{\text{CSRec}} = \frac{1}{N}\sum_N \delta(d_t,\hat{d}_t), \quad \hat{d}_t = \chi_{\geq 1 - \alpha}(z_t),
\end{equation}
where $z_t$ is the model output, $\alpha$ is a pre-defined threshold; $\delta(a,b) = 1$ when $a = b$ and zero otherwise; $\chi_A(x)$ is an indicator function such that it equals to 1 when $x \in A$ and zero otherwise.

Concretely, for a recommended item and ground truth decision $d_t \in \{0,1\}$, CSRec outputs a predicted probability $z_t \in [0,1]$. Then for a pre-defined $\alpha$, we considered the predicted decision $\hat{d}_t$ as $1$ if $z_t \geq 1 - \alpha$ and $0$ otherwise. Then the accuracy is computed by counting the ratio of the succeeded prediction after this conversion.

For baseline models, we convert the model output into users' decisions based on the ranking of the recommended items. Notice that the baseline SR models do not take previous decisions as explicit input. Hence we filtered the interventional data sequence such that only recommended items with positive decisions are preserved. This strategy is similar to the post-processing procedure for FMLPRec \cite{Zhou2022FMLPRec} and BSARec \cite{shin2024BSARec}.

To convert the ranking of the recommended items to users' purchasing decisions, we consider the users' decision to be $1$ whenever the predicted score for recommended items are in the top-$\alpha$-percent $\mathcal{A}_{\alpha}$. We then compute the binary cross entropy and adjusted hit rate by comparing the converted decisions with the ground truth decision. Mathematically, AHR for baselins are defined as follows:
\begin{equation}
    \text{AHR}_{\text{baselines}} = \frac{1}{N}\sum_N \delta(d_t,\hat{d}_t), \quad \hat{d}_t = \chi_{\mathcal{A}_{\alpha}}(\Tilde{z}_t).
\end{equation}

\begin{remark}
    We computed AHR@$\alpha$ for both CSRec and baseline SR models through the discrepancy of their predicted decision after conversion and the ground truth decision. The difference lies in the conversion due to the difference of the objective of CSRec and baseline SR models. Specifically, results for CSRec are mapped to $\{0,1\}$ according to the absolute value of the output because we model the probability; results for baseline SR models are mapped to $\{0,1\}$ depending on the ranking for the inquired item.
\end{remark}

\subsubsection{Parameter settings.} 
For both of the datasets, the input sequences for both interventional or observational data are composed of the last 50 items before the target time index. 
If the sequence length is less than 50, the sequence is padded with zeros.
We train all models with the Adam optimizer with $\beta_1 = 0.9, \beta_2 = 0.999$. The learning rate is set to b/e 0.0005 for GPT4-books and 0.001 for MSSD, and the batch size is set to be 256. For consistency, the item embedding size is set to be 64 for all models.
All parameters used for training baselines and CSRec can be found in the anonymous GitHub repository.

\subsection{Experimental results}

\subsubsection{\textbf{[RQ1]}. The proposed method outperforms baselines on interventional data.} \Cref{tab:scores_books} and \Cref{tab:scores_mssd} show the performance of CSRec and baseline SR models on both interventional and observational data for datasets GPT4-books and MSSD. Here for CSRec, \Cref{tab:scores_books} shows the result of using FMLPRec for approximating the quantity $\mathbb{P}(D_t | S_t, D_{t-1}, P_t)$, and in \Cref{tab:scores_mssd}, we used BSARec to estimate it. We used the adjusted hit rate (AHR) with two different ratios $0.2$ and $0.5$ as evaluation metrics. As a comparison, we also included the results for WSDM winning model (Few-Shot) and integrated it with our framework (CS-Few-Shot) on real-world MSSD, shown in \Cref{tab:scores_mssd}.

We observe that on both datasets, AHR for $\alpha$ at both $0.2$ and $0.5$ has been improved tremendously. 
On GPT4-books dataset, SASRec has the highest score for AHR@0.2, and CSRec approximately triples the score on GPT4-books, meaning that using $\alpha = 0.2$, the accuracy for predicting users' decisions using CSRec could be almost three times than that using baseline SR models. Even with a more relaxed metric AHR@0.5, on GPT4-books, baseline SR models is only doing slightly better than a random guess. 

On real-world MSSD data, CSRec outperforms SOTA SR models by a factor of 1.5 using AHR@0.2, with a higher score on AHR@0.5 as well. Furthermore, when integrated with WSDM winning model, our framework (CS-Few-Shot) achieves the hit rate accuracy as high as 0.6764 using conversion threshold $\alpha = 0.2$. Compared to the original model, our framework is able to improve the accuracy of modeling users' decision by 64\%.

It is because baseline SR models do not take negative decisions explicitly into consideration. Therefore, they neglected the fact that current decision is affected by previous recommendations (either successful or unsuccessful).
This observation highlights the significance of distinguishing users' natural selection processes and their decision making processes under recommendation.


\subsubsection{\textbf{[RQ2]}. The proposed method does not degrade the performance on observational data.}

We compared CSRec and baseline SR models on the observational data for both GPT4-books and MSSD. Results are shown in \Cref{tab:scores_books} and \Cref{tab:scores_mssd} respectively. 

On MSSD dataset (\Cref{tab:scores_mssd}), we observe that compared to the general SOTA SR models, CSRec  achieves higher score for the five metrics out of six except for HR@20. Moreover, when integrating with the winning model, our framework (CS-Few-Shot) is able to maintain its existing level of performance. 

On GPT4-books dataset (\Cref{tab:scores_books}), though CSRec does not improve the baseline SR models on observational data, it also does not cause any degradation on the performance. 
Specifically, on GPT4-books, CSRec achieved a slightly higher score on HR@20 than BSARec, which has the highest score among all baselines. In general, CSRec is able to receive median scores for all metrics. 

Consequently, CSRec is able to improve existing models on interventional data while maintaining their performance on observational data. 
This reveals that CSRec has a wide application in practice.

\subsubsection{\textbf{[RQ3]}. The proposed method universally improves baseline SR models on interventional data.}

We further integrated CSRec with four baseline SR models: SASRec, BERT4Rec, FMLPRec and BSARec, on interventional data for GPT4-books. We show in \Cref{tab:improv} the improvement of incorporating CSRec into the four recommendation models in absolute scores as well as the percentage change. For all four tested baseline models, CSRec is able to universally improve their performance using both metrics. However, the improvement varies depending on the baseline models used. This might be because the integration of each baseline requires its own tuning scheme. It's important to note that we did not adjust the training parameters for the different baseline models.

\input{tex/improvement}

%% file: tex/gpt4_table.tex
\begin{table*}[!h]
\caption{Recommendation performance comparison of different models on GPT4-books dataset. $X_{\text{obs}}$ denotes the observational data and evaluated using HR@$k$ and NDCG@$k$ for $k = 5, 10, 20$. $X_{\text{intv}}$ denotes the interventional data and evaluated using AHR@$\alpha$ for $\alpha = 0.2, 0.5$.  
Boldface denotes the highest score among all models.}
\vspace{-0.7em}
  \label{tab:scores_books}
  \begin{center}
  \begin{tabular}{llccccccc}
    \toprule
     Data Type & Metric & GRU4Rec & Caser & SASRec & BERT4Rec & FMLPRec & BSARec & CSRec \\
    \midrule
     \multirow{6}{*}{$X_{\text{obs}}$} & $\uparrow$ HR@5  & 0.0492 & 0.0489 & \textbf{0.0533} & 0.0528 & 0.0496 & 0.0524 & 0.0513 \\
      & $\uparrow$ HR@10 & 0.1018 & 0.1001 & \textbf{0.1057} & 0.1055 & 0.1019 & 0.1035 & 0.1035\\
      & $\uparrow$ HR@20 & 0.1988 & 0.2014 & 0.2017 & 0.2011 & 0.2008 & \textbf{0.2062} & 0.2070\\
     & $\uparrow$ NDCG@5 & 0.0292 & 0.0285 & 0.0313 & \textbf{0.0320} & 0.0297 & 0.0316 & 0.0297\\
     & $\uparrow$ NDCG@10 & 0.0460 & 0.0448 & 0.0480 & \textbf{0.0488} & 0.0460 & 0.0479 & 0.0463 \\
     & $\uparrow$ NDCG@20 & 0.0701 & 0.0701 & 0.0720 & 0.0726 & 0.0707 & \textbf{0.0736} & 0.0721 \\
     \cmidrule{1-9}
    \multirow{2}{*}{$X_{\text{intv}}$}& $\uparrow$ AHR@0.2 & 0.2431 & 0.2431 & 0.2438 & 0.2430 & 0.2435 & 0.2405 & \textbf{0.932}\\
    & $\uparrow$ AHR@0.5 & 0.4967 & 0.4942 & 0.4981 & 0.5068& 0.5006 & 0.5037 & \textbf{0.932}\\
  \bottomrule
\end{tabular}
\end{center}
\vspace{-0.5em}
\end{table*}

%% file: tex/mssd_table.tex
\begin{table*}[!h]
\caption{Recommendation performance comparison of different models on MSSD dataset. $X_{\text{obs}}$ denotes the observational data and evaluated using HR@$k$ and NDCG@$k$ for $k = 5, 10, 20$. $X_{\text{intv}}$ denotes the interventional data and evaluated using AHR@$\alpha$ for $\alpha = 0.2, 0.5$.  
Boldface denotes the highest score among all models. We compare sota sequential reccomenders and also the winning model, Few-Shot\cite{chang2019sequential} of the WSDM challenge where MSSD is originally released.  Few-Shot has a specialized design that fully considers the properties of music tracks.}
\vspace{-0.5em}
  \label{tab:scores_mssd}
  \begin{center}
  \begin{tabular}{llcccccccccc}
    \toprule
     &  \multicolumn{9}{c}{\textbf{General SOTA Sequential Recommenders}} &  \multicolumn{2}{c}{\textbf{WSDM winning model}}  \\ 
     \cmidrule(lr){2-9} \cmidrule(lr){11-12} 
     Data & Metric & GRU4Rec & Caser & SASRec & BERT4Rec & FMLPRec & BSARec & CSRec & & Few-Shot & CS-Few-Shot  \\
    \cmidrule(lr){1-9} \cmidrule(lr){11-12} 
     \multirow{6}{*}{$X_{\text{obs}}$} & $\uparrow$ HR@5 & 0.1045 & 0.0963 & 0.0845 & 0.0219 & 0.1487 & 0.1736  &\textbf{0.1745} & & \textbf{0.1865} & 0.1798\\ 
     & $\uparrow$ HR@10 & 0.1311 & 0.1012 & 0.1082 & 0.0407 & 0.1952 & 0.1901  &\textbf{0.2007} & & 0.2235 & \textbf{0.2364}\\
     & $\uparrow$ HR@20 & 0.1944 & \textbf{0.3031} & 0.2110 & 0.0769 & 0.2747 & 0.2878  & 0.2871 & & 0.3712 & \textbf{0.3731}\\
     & $\uparrow$ NDCG@5 & 0.0148 & 0.0286 & 0.0429 & 0.0341 & 0.0745 & 0.1039 & \textbf{0.1041} & & \textbf{0.1336} & 0.1325\\
     & $\uparrow$ NDCG@10 & 0.0312 & 0.0457 & 0.0981 & 0.0771 & 0.1031 & 0.1487 &\textbf{0.2117} & &\textbf{0.1674} & 0.1572\\
     & $\uparrow$ NDCG@20 & 0.0773 & 0.0717 & 0.1227 & 0.0985 & 0.1592 & 0.2793 &\textbf{0.2801} & & 0.3412 & \textbf{0.3435}\\
     \cmidrule{1-12}
     \multirow{2}{*}{$X_{\text{intv}}$} & $\uparrow$ AHR@0.2 & 0.1117 & 0.1461 & 0.1926 &  0.1468 & 0.1412 & 0.1692 & \textbf{0.3509} && 0.4127 & \textbf{0.6764} \\
     & $\uparrow$ AHR@0.5 & 0.2403 & 0.2467 & 0.2956 & 0.3401 & 0.3009 & 0.4323  &\textbf{0.5371} & &0.5061 & \textbf{0.7418}\\
  \bottomrule
\end{tabular}
\end{center}
\vspace{-1em}
\end{table*}

%% file: tex/improvement.tex
\begin{table}[!h]
\caption{ Improvement of CSRec integrated with baselines SASRec, BERT4Rec, FMLPRec and BSARec on GPT4-books using AHR@0.2 and AHR@0.5. Each column $\delta_{\text{model}}$ represents the improvement in scores after applying CSRec on each model with the percentage change in the parenthesis. }
\vspace{-0.5em}
  \label{tab:improv}
  \begin{center}
  \footnotesize
  \begin{tabular}{ccccc}
    \toprule
     Metric & $\delta_\text{SASRec}$ & $\delta_\text{BERT4Rec}$ & $\delta_\text{FMLPRec}$ & $\delta_\text{BSARec}$ \\
    \midrule
    $\uparrow$ AHR@0.2 & 0.1476 (60.5\%) & 0.0616 (25.5\%) & 0.6885 (283\%) & 0.6845 (285\%)\\
    $\uparrow$ AHR@0.5 & 0.4195 (84.0\%) & 0.4085 (80.6\%) &0.4314 (86.2\%)& 0.4283 (85.2\%)\\
  \bottomrule
\end{tabular}
\end{center}
\end{table}

%% file: sections/Related_Work.tex
\section{Related Work}
\noindent \textbf{Sequential Recommender Systems}
Sequential recommender systems (SRSs) learn the user-item interactions (e.g., view or purchase items on an online shopping platform) by modelling the dependencies in a sequence. Contrary to the traditional recommender systems which focus more on static learning through content-based and collaborative filtering\cite{wang2019sequential}, SRSs treat the interaction in a more dynamic way and hope to capture the inherent logic between the current and recent preference of a user. Traditional works formulate the task of sequential prediction as a first-order or high-order Markov Chain\cite{he2017translation}, whereas more recent works make use of the rapid development of the deep neural networks and propose RNN-based\cite{donkers2017sequential} and, more recently, attention-based\cite{kang2018self, shin2024attentive} frameworks to capture the temporal features. \cite{qiu2022contrastive, du2023frequency} applies contrastive learning strategies to help the learning of the sequential dependencies, \cite{zhou2023attention} adjusts unreliable attention weights for a more robust generalization. Despite the great success, there are two fundamental challenges that remain unresolved. However, the above works equate preference to decisions. However, user preference towards the item is only a partial reason for a purchase decision. For example, a user is highly unlikely to purchase another refrigerator no matter how they like it. Such interactive relationship is not fully captured or even modeled in existing works. In our work, we propose a framework that accurately characterizes the causal relationship between the recommendation from the system, the user's past purchasing behaviour and the user's preference, obtaining universally better performance.

\noindent \textbf{Causal Inference in Recommendation.}
 Causal inference methods are widely used in estimating the causal effect\cite{pearl2012calculus}, understanding the compound causal relationship among various factors\cite{liu2023cat,wang2021desiderata}, injecting external knowledge to the model\cite{liu2024c}, and isolating the causal relationship of certain variables. Causal inference is to estimate potential outcomes when treatment is applied\cite{yao2021survey,liu2024large}. It originated from the scenario of prescribing medicine to patients. The recommendation is naturally a causal inference task when treating each item as a potential treatment. The goal is to predict the user's preference towards these treatments(i.e., the recommended items). 
There is a line of works framing recommendation as a causal inference task \cite{gao2024causal}. \cite{schnabel2016recommendations} show how estimating the quality of a recommendation system can be approached with Inverse Propensity-reWeighting(IPW) techniques commonly used in causal inference, and the proposed method is based on matrix factorization.\cite{wang2020causal} is also based on traditional matrix factorization. They make separate estimations to correct unobserved confounders (the factor that affects both treatment assignment and potential outcome). \cite{xu2021causal} propose a general framework for modelling causality in collaborative filtering and recommendation. \cite{zhang2021causerec} applies contrastive learning to solve the counterfactual reasoning problem. \cite{chen2021autodebias} derives a universal strategy for fixing various biases. Compared to traditional non-causal counterparts, causal recommender systems obtained successes in user-item interaction prediction and de-bias observational training data. In this paper, we propose the first causal framework for quantifying sequential recommendation and introduce a constraint that maintains the casual relationship in subsequent purchases in the modelling.

%% file: sections/Conclusion.tex
\section{Conclusion}
\label{sec:conclusion}
\noindent \textbf{Conclusion.} In this paper, we aim to address the question: How do recommender systems impact users' decisions in the context of causal inference? 
By distinguishing between user natural selection process and their decisions making under recommendation, we propose a framework to model the causal relationship between user's decision and various factors including their natural preference and their previous decision towards recommendations.
We introduce CSRec, a general framework with straightforward implementation that can easily adapted to SOTA sequential recommenders. Experimental results show that our framework is capable of improving the accuracy of predicting users' decision under recommendations without causing any degradation on non-interventional setting. 
Lastly, we further propose to isolate the system's influence on users through treatment effect estimation, which opens new possibilities in designing personalized recommender systems.

%% file: sections/app3_prelim.tex
\appendix
\newpage

\section{Do-Calculus}
\label{app: do-calculus}


\begin{rules}[Insertion/deletion of observations] 
\label{thm:rule1}
\begin{equation}
    P(y|do(x), z, w) = P(y|do(x), w) \quad \text{if} \quad (Y \indep Z | X, W)_{G_{\overline{X}}}
\end{equation}
\end{rules}

\begin{rules}[Action/observation exchange] 
\label{thm:rule2}

\begin{equation}
    P(y|do(x), do(z), w) = P(y|do(x), z, w) \quad \text{if} \quad (Y \indep Z | X, W)_{G_{\overline{X}\underline{Z}}}
\end{equation}
\end{rules}

\begin{rules}[Insertion/deletion of actions] 
\label{thm:rule3}

\begin{equation}
       P(y|do(x), do(z), w) = P(y|do(x), w)  \quad  \text{if} \quad (Y \indep Z | X, W)_{G_{\overline{X}\overline{Z(W)}}}
\end{equation}
\end{rules}

where $G_{\overline{X}}$ is the graph with all incoming edges to $X$ being removed,  $G_{\underline{W}}$ is the graph with all outcoming edges to $W$ being removed, and $Z(W)$ is the set of $Z$-nodes that are not ancestors of any $W$-node.

Intuitively, \Cref{thm:rule1} states when an observant can be omitted in estimating the interventional distribution, \Cref{thm:rule2} illustrates under what condition, the interventional distribution can be estimated using the observational dataset, and \Cref{thm:rule3} decides when we can ignore an intervention.

%% file: sections/app1_thm.tex


\section{Proof for \Cref{thm:prob_dist}}
\label{app:proof_thm}
\begin{proof}
    Suppose that the user's decision $D_t$ follows the causal relationship in \Cref{fig:causal-graph}, then we can write
\begin{multline}
    \mathbb{P}(D_t | \text{do}(S_t, S_{t-1},...,S_1), P_t) = \sum_{D_{t-1}} \mathbb{P}(D_t | \text{do}(S_t, S_{t-1},...,S_1), P_t, D_{t-1}) \cdot \\
    \mathbb{P}(D_{t-1}|\text{do}(S_t, S_{t-1},...,S_1), P_t)
\end{multline}
From the figure we see that there's no causal influence of $S_t$ on the decision $D_{t-1}$, hence we can write 
\begin{equation*}
    \mathbb{P}(D_{t-1}|\text{do}(S_{t-1},...,S_1), P_t)
\end{equation*}
We know that $D_t$ and $S_t$ satisfies 
$$(D_t \indep S_t | S_{t-1},..., S_1, P_t, D_{t-1})_{G_{\overline{S_{t-1},..., S_1}, \underline{P_t,D_{t-1}}}}.$$
Hence, using \Cref{thm:rule2}, we can obtain 
\begin{equation*}
    \mathbb{P}(D_t | \text{do}(S_t, S_{t-1},...,S_1), P_t, D_{t-1}) = \mathbb{P}(D_t | S_t, \text{do}(S_{t-1},...,S_1), P_t, D_{t-1}).
\end{equation*}
From the causal graph, we can also see that 
$$(D_t \indep \{S_{t-1},...,S_1\}|S_2, D_{t-1}) G_{\overline{\{S_{t-1},...,S_1\}}(S_t, D_{t-1})}.$$
Therefore, \Cref{thm:rule3} leads to 
\begin{equation*}
    \mathbb{P}(D_t | S_t, \text{do}(S_{t-1},...,S_1), P_t, D_{t-1}) = \mathbb{P}(D_t | S_t, P_t, D_{t-1})
\end{equation*}
Combining the results for the two component we proved the result for \Cref{thm:prob_dist}. 
\end{proof}

%% file: sections/app2_exp.tex
\section{Experimental Details}
\label{app: exp-details}

\subsection{Spotify: Sequential Music Streaming Challenge}
\label{subsec:spotify}
\MSSD is first published at the WSDM Cup 2019 Spotify Sequential Skip Prediction Challenge \cite{brost2019music}. It has approximately 160 million listening sessions, associated user actions, and 3.7 million tracks with acoustic features. The task is to predict whether a particular user will skip individual tracks encountered in a listening session. The sessions in this dataset are sampled from various sources such as radio, personalized recommendations, users' own collections, expertly curated playlists, and contextual but non-personalized recommendations. Records of listening sessions from users' own collection approximate the observational dataset while 
the \emph{personalized recommendation} source serves as an ideal interventional dataset for studying the user-system dynamics from a causal perspective: users who listen to personalized recommendations do not have direct control over what they will hear. In other words, the exposure is exogenously determined by the RecSys, which reflects Spotify's belief in the users' preferences. In the remainder of the paper, we focus on the listening sessions collected from such a personalized recommendation setting as the interventional data, and the user's own collection as the observational data.

Following the causal perspective of recommendation, the treatment is considered effective if the user does not skip it and the treatment is presented as a ranked list of the songs selected for the users. Therefore, we define two tasks of interest on \MSSD: skip prediction and track ranking. Skip prediction measures the individual treatment effect of each track for a certain user, while we measure the relative effectiveness among all treatments when ranking the tracks. Skip prediction is equivalent to estimating treatment effect, and the track ranking corresponds to selecting the most effective treatments. To preserve user privacy, Spotify randomly cut the listening sessions in MSSD from longer sessions so that a single session contains no more than 20 songs. 

\begin{table*}[!h]
\caption{Results for CSRec integrated with baselines SASRec, BERT4Rec, FMLPRec and BSARec on GPT4-books using AHR@0.2 and AHR@0.5. CSRec + <model> represents the performance of CSRec integrated with baseline <model>.}
\vspace{-0.5em}
  \label{tab:improv}
  \begin{center}
  \begin{tabular}{lcccccccc}
    \toprule
     Metric & SASRec & CSRec + SASRec & BERT4Rec & CSRec + BERT4Rec & FMLPRec & CSRec + FMLPRec & BSARec & CSRec + BSARec \\
    \midrule
    $\uparrow$ AHR@20 & 0.2438 & 0.3914 & 0.2430 & 0.3046 & 0.2435 & 0.932 & 0.2405 & 0.925\\
    $\uparrow$ AHR@50 & 0.4981 & 0.9176 & 0.5068 & 0.9153 & 0.5006 & 0.932 & 0.5037 & 0.932\\
  \bottomrule
\end{tabular}
\end{center}
\end{table*}

\subsection{GPT4-books}
\label{app: data-generation}
Given the limited access to interventional data, where the recommender system fully controls the exposure. 
Previous work has explored the potential of using a large language model agent to generate synthetic data for downstream tasks \cite{liu2024bestpracticeslessonslearned, zhou2024explorespuriouscorrelationsconcept, zhou2024multi,chen2024genqa,chen2024automated,zhou2024calibrated}.
To resolve the data shortage, we also created a synthetic dataset using GPT-4 as agents to mimic users' reactions to certain recommendations.
Specifically, we use GPT4-turbo as the agent that mimics users with various tastes of books. 

\noindent\textbf{Preparing the item list.} We first sampled a list of book names from the Amazon Book dataset\cite{ni2019justifying}. We use books as items to recommend because GPT4 has had prior information towards these items. We then asked the GPT4 about the basic information about the books. The reason is that we want to validate the GPT4 knows the books and is be able to serve as potential readers. Specifically we validate that by asking the GPT about basic information and keywords about the book names that we provided and comparing them with what we have in the Amazon Book dataset.

\noindent\textbf{Synthetic users.} The second step is to create synthetic users. We refer to the scale of similar datasets used by existing baselines\cite{hidasi2016GRU4Rec,tang2018caser,shin2024BSARec,sun2019bert4rec,Zhou2022FMLPRec} and define 10000 users. Each user is defined by a probability distribution of preference across different genres. For example, we describe a user as who’s preference towards history, fiction, novels, and romance is $25\%,  25\%,  25\%,  25\%$ where the probabilities are randomly generated.

\noindent\textbf{Synthetic observational/interventional dataset.}
For each user, we asked GPT4 to generate the synthetic purchase records of various length. Specifically, we collect the top 100 popular books for each genre, and randomly set a length $l$ for each user. We then asked the GPT4 to generate a sequence of length $l$,which is the most possible purchase records for this user, among the list of books. For the interventional dataset, instead of asking GPT4 what is the most possible purchase records, we generate random sequences of recommendations, and asked the agent whether the recommended book will be purchased, and the probability of such a decision. Timestamp information is also provided to GPT4. 

\subsection{Generalization of the causal graph used in this paper.}
In this paper, we propose a novel approach to model the recommendation process from a causal perspective. We begin by constructing a causal graph that represents the recommendation scenario of interest. This graph allows us to identify how a user's decisions are influenced by various factors within the system. 
Rather than considering the sequential recommendation as a next-item prediction task,
our approach focuses on understanding how a user's decisions can be positively influenced by specific factors.

As illustrated in \cref{fig:causal-graph}, the recommender system is conceptualized as an intervention that impacts the user's decision-making process. This causal graph is versatile and applicable to both intervention-based and observational scenarios. In the former, the recommender system acts as the direct intervention, while in the latter, the user's decisions are assumed to be influenced by unobserved sources. Thus, our model provides a framework for understanding the effects of both observed and unobserved interventions on user behavior.

\subsection{Additional Results.}
See \Cref{tab:improv}